\documentstyle[aps]{revtex}
\begin{document}

\title{Coherent states and uncertainty relations \thanks{
Publised in "Quantum Field Theory, Quantum Mechanics and Quantum Optics,
Pt.I: Symmetries and Algebraic Structure in
Physics", Edited by V. V. Dodonov and V. I. Man'ko, 1991,
Nova Science Publishers, Inc., Singapore, pp. 247-249.}} 
\author{Horia Scutaru}
\address{Department of Theoretical Physics, Institute of Atomic Physics,
\\ PO Box MG-6\\ Bucharest-Magurele, Romania}
\maketitle
\begin{abstract}
A sharp estimation of the $L^p$-norms of some matrix 
coefficients of the square integrable representations
is conjectured. The conjecture can be proved for integer
values of $p$ using a result of J. Burbea.
\end{abstract}

In an unpublished paper \cite{scut} we have obtained the
following sharp estimation of the $L^p$-norms of the matrix 
coefficieents of the square integrable representations
$U_{k}(x,y,t)$, with $k \in {\bf R}$, 
of the Heisenberg group (which is defined as
the manifold ${\bf R}^3$ with the group multiplication
given by the rule : $(x,y,t)(x^{'},y^{'},t^{'})=(x+x^{'},
y+y^{'},t+t^{'}+{1 \over 2}(x{'}y-xy^{'}))$):
\begin{equation}
(|k|(2 \pi)^{-1} \int_{{\bf R}}|(h,U_{k}(x,y,t)f)|^{p} dxdy)^{{1
\over p}} \leq ({2 \over p})^{{1 \over p}}||h|| ||f||
\end{equation}
for all $h,f \in L^2({\bf R})$ and where the equality is
attained iff $h$ and $f$ are Glauber coherent states.
This result was quoted in \cite{grab} in connection with the
uncertainty relations. 
In paper \cite{scut} we have considered, also, the following
immediate question raised by the above result: {\it do there exist
such sharp estimations for the square integrable representations
of other locally compact unimodular Lie groups ?} 
Let us denote by $G$ such a group and let  $U$ denote a unitary
representation of $G$ in a Hilbert space $H$. We suppose that
this representation is a representation with coherent states, 
i.e. there is a subgroup $S$ of $G$, a character $e: S \rightarrow
{\bf T}$ of $S$, and a vector $f_{0}$ of $H$ such that:
\begin{equation}
U(s)f_{0}=e(s)f_{0},~~~s \in S,
\end{equation}
and
\begin{equation}
\int_{G/S}|(h,U(g)f_{0})|^2 d \dot g  <  \infty ,~~~h \in H.
\end{equation}
Here $\dot g$ is the Haar measure on $G/S$. Then, it is well known
that there exists a real nonvanishing number $dim(U)$ so that:
\begin{equation}
dim(U) \int_{G/S}|(h,U(g)f_{0})|^2 d \dot g = ||h||^2 ||f_{0}||^2,~~~
h \in H.
\end{equation} 
For such representations a straigthforward generalization of the
result (1) is given by the following conjecture:
\vskip 0.5cm
{\bf CONJECTURE}. {\it For any real $p \geq 2$ and for any $h
\in H$ there exists a constant $C(p) \leq 1$ with $C(2)=1$ and
such that:
\begin{equation}
\left(dim(U) \int_{G/S}|(h,U(g)f_{0})|^p d \dot g \right)^{{1 \over p}} 
=C(p) ||h|| ||f_{0}||,~~~
h \in H.
\end{equation}
and where the equality is attained iff $h$ is a coherent state,
i.e. iff $h=cU(g)f_{0}$, $g \in G$, $c \in {\bf C}$.}
\vskip 0.5cm
In the following we shall discuss the relevance of this
conjecture for the uncertainty relations and we shall prove it
when $p$ is any even natural number and $G$ is one of the
following groups: the Heisenberg group, the group $SU(2)$
and the group $SU(1,1)$.

From (4) it follows that the Hilbert space $H$ is embedded
in the Hilbert space $L^2(G/S,dim(U)d \dot g)$ as a subspace
with reproducing kernel. Hence for any vector $h \in H$ with
$||h||=1$ we can define the wave function $(h,U(g)f_{0})$
on $G/S$ and a probability distribution $P(\dot g)=
|(h,U(g)f_{0})|^2$ on $G/S$. Then it is evident that the
left hand side of (5) can be considered as a measure of the
extent to which the above defined wave function and the
corresponding probability distribution are peaked on $G/S$.
From (5) it follows that the probability distribution
$P(\dot g)$ on $G/S$ cannot be arbitrarily peaked on the
phase space $G/S$. Hence (5) is an uncertainty relation
for the wave function defined on the phase space and the
most peaked wave functions are those associated with
the coherent states.

When $G$ is one of the three particular cases, enumerated above,
the phase space $G/S$ will be denoted by $\Delta$ and is, respectively,
the complex plane ${\bf C}$, the unit disc and the Riemann sphere.
The irreducible representations with coherent states on $\Delta$
of the first and last group can be parametrrized by a positive
real number $\beta$ and the corresponding generalized dimension
$dim(\beta)$ is defined by: $dim(\beta)=\beta$ for the
Heisenberg group and $dim(\beta)=\beta-1$ for the group $SU(1,1)$.
For the group $SU(2)$ the parameter $\beta$ takes only integer
and semiinteger values and the generalized dimension $dim(\beta)$ is
equal with the ordinary one which is given by the well known formula:
$dim(\beta)=2 \beta +1$. The invariant measure $d \dot g$ on $\Delta$
is given by $d \dot g=\pi^{-1}dxdy$ for the Heisenberg group and
by $d \dot g =\pi^{-1}(1 \pm |z|^2)^{-2}dxdy$, $(z=x+iy)$, for the
groups $SU(2)$ and $SU(1,1)$ respectively. In all three cases the
Hilbert space $H_{\beta}$ is unitarily equivalent with the space
of holomorphic functions on $\Delta$ with the norm defined by:
\begin{equation}
||f||_{\beta}=dim(\beta)\int_{\Delta}|f(z)|^2 k_{\beta}(|z|^2)^{-1}
d \dot g,
\end{equation}
where $k_{\beta}(z)=\exp(\beta z)$ for the Heisenberg group,
$k_{\beta}(z)=(1-z)^{-\beta}$ for the group $SU(1,1)$, and
$k_{\beta}(z)=(1+z)^{2\beta}$ for the group $SU(2)$.
The reproducing kernel is given by $k_{\beta}(z)$
In order to prove the conjecture we shall use the following
theorem proved in \cite{burb}:
\vskip 0.5cm
{\bf THEOREM}. {\it If $f \in H_{\beta}$ and $h \in H_{\beta^{'}}$,
then $fh \in H_{\beta + \beta^{'}}$ and
\begin{equation}
||fh||_{\beta+\beta^{'}} \leq ||f||_{\beta} ||h||_{\beta^{'}}.
\end{equation}
The equality is attained iff either $fh=0$ or $f$ and $h$ are of
the form $f=c_{1}k_{\beta}(\bar w z)$, 
$h=c_{2}k_{\beta^{'}}(\bar w z)$ for
some $w \in \Delta$ and some nonzero complex constants $c_{1}$ and
$c_{2}$.
}
\vskip 0.5cm
As a corollary one obtain \cite{burb} for any natural number
$n$:
\begin{equation}
||f^n||_{n \beta} \leq ||f||_{\beta}^{n}
\end{equation}
with the equality attained either for $f=0$ or when
$f=ck_{\beta}(\bar w z)$ for some nonzero complex constant $c$.

From (8) it follows for the probability distribution on
$\Delta$ given by $P_{\beta}(z)=|f(z)|^2k_{\beta}(|z|^2)^{-1}$
that:
\begin{equation}
\int_{\Delta}P_{\beta}(z)^n d \dot g \leq {dim(\beta) \over
dim(n \beta)}
\end{equation}
with the equality for $P_{\beta}(z)=k_{\beta}(\bar w z)^2
k_{\beta}(|z|^2)^{-1}k_{\beta}(|w|^2)^{-1}$.
Hence, the most concentrated wavefunction defined on the
phase space $\Delta$ is the coherent state
$ k_{\beta}(\bar w z)
k_{\beta}(|z|^2)^{-{1 \over 2}}
k_{\beta}(|w|^2)^{-{1 \over 2}}$.


\begin{references}

\bibitem{scut}
H. Scutaru, Sharp inequalities for $L^{p}$-norms of matrix
coefficients of square integrable representations of
the Heisenberg group and of $L^{p}$-norms of Wigner
functions, preprint FT-167-79, Central Institute of Physics,
Bucharest, 1979.
\bibitem{grab}
M. Grabowski, Reports on Mathematical Physics {\bf 20} 153 (1984).
\bibitem{burb}
J. Burbea, Illinois J. Math. {\bf 27} 130 (1983).
\end{references}
\end{document}